\documentstyle[11pt,epsfig]{article}
\newcommand{\LP}{log-periodic oscillations}
\newcommand{\be}{\begin{equation}}
\newcommand{\ee}{\end{equation}}
\newcommand{\rp}{\right)}
\newcommand{\lp}{\left(}
\newcommand{\bea}{\begin{eqnarray}}
\newcommand{\eea}{\end{eqnarray}}

\begin{document}

\title{Financial ``Anti-Bubbles'': \\ Log-Periodicity in Gold and Nikkei
collapses}
\thispagestyle{empty}

\author{A. Johansen$^1$ and D. Sornette$^{1,2,3}$\\
$^1$ Institute of Geophysics and
Planetary Physics\\ University of California, Los Angeles, California 90095\\
$^2$ Department of Earth and Space Science\\
University of California, Los Angeles, California 90095\\
$^3$ Laboratoire de Physique de la Mati\`{e}re Condens\'{e}e\\ CNRS UMR6622 and
Universit\'{e} de Nice-Sophia Antipolis\\ B.P. 71, Parc
Valrose, 06108 Nice Cedex 2, France}

\thispagestyle{empty}

\maketitle

\vskip 1cm

\begin{abstract}

We propose that imitation between traders and their herding behaviour not
only lead to speculative bubbles with accelerating over-valuations of
financial markets possibly followed by crashes,
but also to ``anti-bubbles'' with decelerating market devaluations
following all-time highs.
For this, we propose a simple market dynamics model in which the demand
decreases
slowly with barriers that progressively quench in, leading to a
power law decay of the market price decorated by decelerating log-periodic
oscillations.
We document this behaviour on the Japanese Nikkei stock index from 1990 to
present
and on the Gold future prices after 1980, both after their all-time highs.
We perform simultaneously a parametric and non-parametric analysis that are
fully consistent
with each other. We
extend the parametric approach to the next order of perturbation, comparing
the log-periodic fits with one, two and three log-frequencies, the latter one
providing a prediction for the general trend in the coming years. The
non-parametric power
spectrum analysis
shows the existence of log-periodicity with high statistical significance, with
a prefered scale ratio of $\lambda \approx 3.5$ for the Nikkei index
$\lambda \approx 1.9$ for the Gold future prices,
comparable to the values obtained for speculative bubbles leading to crashes.

\end{abstract}

\vspace{4cm}

{\bf Pacs numbers: 01.75+m ; 02.50-r ; 89.90+n}

\newpage
\pagenumbering{arabic}

\section{Introduction}

Since our first suggestion of the existence of specific
log-periodic price patterns
associated to the speculative accelerating power law
bubble leading to the worldwide Oct. 1987 stock market crash
\cite{stockcrash1},  more evidences have accumulated [1-11]. They
include the Oct. 1929
US crash \cite{Feigenbaum,stockcrash2}, the May 1962 US market
correction of $15\%$ discovered in a
systematic blind-testing procedure \cite{jls,thesis}, the Hong-Kong Oct. 1997
crash \cite{jls,thesis},
the Oct. 1997 US market correction [6-9] (see also the
prediction communicated on sept. 17, 1997
to the French office for the  protection of
proprietary softwares and inventions under number registration 94781 and
footnote 13 of
Ref.\cite{StauSor}) and the Aug. 1998 US market correction of $19\%$
\cite{risk}.

Recently, the same log-periodic signatures with similar prefered scaling
ratios and power law exponents have been found to describe the speculative
behaviour in the Reagan's years of the US dollar against other major
currencies, such as the German Mark and the
Swiss Franc up to the maximum of the bubble in March 1985 \cite{risk}.
It is remarkable to find that very similar structures describe both the
behaviour of
entire markets quantified by their indices and the behaviour of the Foreign
exchange (Forex). The two are generally believed to obey different dynamics
\cite{Chopard}, since on the Forex only a few major banks perform the
trading and only a few currencies (the US\$ against the DEM now replaced by
the Euro and the YEN) account for most (more than $80\%$) of the trading.

This strengthens the concept that these structure are characteristic of
very fundamental and robust properties of financial markets, which we have
proposed to be the herding behaviour of traders \cite{jls,risk,Physica2}.
Furthermore, it supports the view according to which the stock markets
present self-organising properties similar to those of microscopic models
in statistical physics, with the potential for cooperative behaviour
\cite{AAP88} leading to critical points \cite{manifest,Moneysex} (for
criticism,
see \cite{Laloux}).

In addition, these results suggest that the prefered scale ratio $
\lambda$ of the underlying discrete scale invariance \cite{Review} of
the market dynamics
is remarkably robust with respect to changes in {\it what} is traded
and {\it when} it is traded. These results are at odds with the standard
random walk null-hypothesis and the efficient market hypothesis.
This hypothesis leads to the conclusion
that stock and currency markets are unpredictable.

The question we address here is whether the cooperative herding behaviour of
traders might also produce market evolutions that are symmetric
to the accelerating speculative
bubbles often ending in crashes. This symmetry is performed with respect to
a time inversion
around a critical time $t_c$ such that $t_c-t$ for $t<t_c$ is changed into
$t-t_c$ for $t>t_c$.
This symmetry implies to look at {\it decelerating} devaluations instead of
accelerating valuations. A related observation has been reported in our first
paper \cite{stockcrash1} showing that the implied volatility of traded
options, which is a
measure of the perceived market risk, has relaxed {\it after} the Oct. 1987
crash to its
long-term value, from a maximum at the time of the crash,
 over a time-scale of about a year, according to a decaying power law with
decelerating log-periodic oscillations. It is this type of
behaviour that
we document now but for real prices.

We show that there seems to exist critical times $t_c$
at which the market culminates, with either a power law increase with
accelerating
log-periodic oscillations or a power law decrease with decelerating
log-periodic
oscillations. We have not found a market for which both phenomena are
simultaneously observed for the same $t_c$. The main reason is that
accelerating
markets with log-periodicity almost always end-up in a crash, a market
rupture that
thus breaks down the symmetry ($t_c-t$ for $t<t_c$ into  $t-t_c$ for
$t>t_c$). The main
message that we draw here is that herding behaviour can progressively occur and
strengthen itself in ``bearish'' (decreasing) market phases, even if the
preceding
``bullish'' phase ending at $t_c$ was not characterised by a
strengthening imitation. The symmetry
is thus statistical or global and holds in the ensemble
rather than for each single case individually. The breakdown of local symmetry
around the critical point $t_c$ is not unknown in thermodynamic phase
transitions. Let
us mention the $\lambda-$transition in $^4 He$, so named because of the
asymmetric
shape of the specific heat around $T_{\lambda}$ with a more abrupt decay above
$T_{\lambda}$ than below \cite{Goldenfeld}
qualitatively similar to a market price time series
around a crash.

The organisation of the paper is as follows. We first present the Landau
expansion
and its extension to third-order. The three derived log-periodic formulas
are used to fit
the Japanese Nikkei index drop since 1990. They agree very well in the
domains where
they overlap, not only in their shape but also in the coherence of their
parameters.
The new third-order formula gives an excellent fit up to the end
of 1998 and provides a prediction for the general trend of 1999 and 2000.
Further empirical evidence is presented on the Gold devaluation a few months
after its peak in 1980 over a period of more than two years. From
parametric fits with log-periodic formulas and a non-parametric log-periodic
power spectrum analysis,
we find strong evidence of structures very similar to those found previously
prior to crashes. We then propose a simple dynamical model
of imitative ``bearish'' market, which using the empirical tests, lead to the
prediction that
the relative strength of the optimistic phase in an otherwise pessimistic
bearish market is stronger than the occasional pessimism
in an otherwise speculative bullish bubble.

\section{Data analysis: Nikkei and Gold}

\subsection{Landau expansion and log-periodic formulas}

A very powerful and general tool used in the studies of critical phase
transitions is
that of Landau expansions. They amount to assuming some functional relationship
$F\lp \tau \rp$ between the relevant observable $F$ and the corresponding
governing
parameter $\tau$. With respect to the financial markets, the observable can
be the
price (or some measure of it) and the governing parameter is time (or some
measure of it). In Ref.\cite{stockcrash2}, we have proposed a
general form for an evolution equation for the complex variable $F$,
obtained by
expanding $F$ around $0$ as
\be \label{Landau}
\frac{d\log F\lp \tau \rp}{d\log \tau}=\alpha F\lp \tau \rp +\beta  |F\lp
\tau \rp|^2
F\lp \tau \rp \ldots ,
\ee
where in general the coefficients may be complex. In this expansion,
only terms that ensure that the equation remains
invariant with respect to a change of phase $\phi \to \phi + C$ where
\be \label{qmqmmqm}
F \equiv B e^{i\phi}
\ee
are allowed since a phase translation corresponds to a change of time units
\cite{stockcrash2}.

A first (keeping only the
$\alpha F\lp \tau \rp$ term in the r.h.s. of (\ref{Landau})) and second order
solution (keeping the two first terms in the r.h.s. of (\ref{Landau}))
gives the following two equations for the price (or some measure of it)
evolution \cite{stockcrash2}
\bea \label{1feq}
p\lp t\rp &\approx& A + B \tau ^{\alpha}
+C \tau^\alpha \cos\left[ \omega\log \lp \tau \rp +\phi \right]
\\ \label{2feq}
p(t) &\approx& A' + \frac{\tau^\alpha}{\sqrt{1+\left(\frac{\tau}
{\Delta_t}\right)^{2\alpha}}} \left\{B'+ C'\cos\left[\omega\log \tau +
\frac{\Delta_\omega}{2\alpha}\left(1+\left(\frac{\tau}
{\Delta_t}\right)^{2\alpha}\right)+\phi'\right]\right\},
\eea
where $\tau = t_c - t$. The log-periodic frequency $\omega$ is related
to the prefered scaling ratio by
\be
\ln \lambda = {2\pi \over \omega}~.
\ee
These two equations (\ref{1feq},\ref{2feq})
describe the evolution of the price {\em prior} to a time
$t_c$, where a large crash may occur, i.e., $t<t_c$. Equation \ref{1feq}
have been found to describe the price evolution up to 3 years prior to large
crashes \cite{risk} and eq. \ref{2feq} up to 8 years \cite{stockcrash2}.
Observe that
eq.(\ref{2feq}) predicts the transition from the log-frequency $\omega$ close
to $t_c$ to $\omega + \Delta_\omega$ far from $t_c$ (i.e. for $\tau >
\Delta_t$).

From the point of view of critical points and within Landau
expansions, there is nothing special about the direction of the ``flow'' or
equivalently the sign of $\tau$ in eq. \ref{Landau}. We thus propose to use
equations (\ref{1feq},\ref{2feq}) replacing $\tau$ with $-\tau$ (or
$t_c-t$ with $t-t_c$), where now $t > t_c$. This novel situation corresponds
to the case of a (power law) decaying price, i.e.,
an ``anti-`bubble'' or, in financial colloquial terms, a ``bearish'' phase.

We will use below an extension of (\ref{1feq},\ref{2feq}) obtained from the
Landau equation (\ref{Landau}) expanded up to the next order
\be
\frac{d\log F\lp \tau \rp}{d\log \tau}=\alpha F\lp \tau \rp +\beta
|F\lp \tau \rp|^2 F\lp \tau \rp  + \gamma |F\lp \tau \rp|^4 F\lp \tau
\rp\ldots ~.
\label{jfmqmqm}
\ee

Inserting (\ref{qmqmmqm}) into (\ref{jfmqmqm}) gives
\bea \label{firstB}
{d B \over d\log \tau} &=& a_1 B + b_1 |B|^2 B + c_1 |B|^4 B~,\\
\label{secondphi}
{d \phi \over d\log \tau} &=& a_2  + b_2 |B|^2  + c_2 |B|^4 ~,
\eea
where
\bea \label{qlql}
\alpha &=& a_1 + i a_2~,\\ \label{jfqmq}
\beta &=& b_1 + i b_2~,\\ \label{jfqqqqmq}
\alpha &=& c_1 + i c_2~. \label{jfqmggdq}
\eea
The solution of (\ref{firstB}) is given by the implicit equation
\be
{\biggl[{B^2 - B_+^2 \over B^2}\biggl]^{B_-^2} \over \biggl[{B^2 - B_-^2
\over B^2}\biggl]^{B_+^2}}
= \biggl({\tau \over \tau_0}\biggl)^{2c_1 B_-^2 B_+^2 (B_+^2 -B_-^2)}~,
\label{ujfmqmqm}
\ee
where
\be
B_{\pm}^2 = {1 \over 2c_1} \biggl( -b_1  \pm \sqrt{b_1^2 - 4a_1c_1} \biggl)~.
\ee
Certain conditions must hold for a solution to exist, in particular $b_1/c_1<0$
to ensure that
the first non-linear correction leads to another stable fixed point and
$b_1^2 \geq 4a_1c_1$ for $B$ to be real.

Depending upon the values of the parameters, the solution $B(\tau)$ of
(\ref{ujfmqmqm}) can take many different functional shapes. This leads
in general to the problem of model misspecification. Here, for the purpose
of simplicity, we choose the set of parameters
such that the solution can be approximated by
$$
p(t) \approx A' + \frac{\tau^\alpha}{\sqrt{1+\left(\frac{\tau}
{\Delta_t}\right)^{2\alpha} + \left(\frac{\tau}{\Delta_t'}\right)^{4\alpha}}}
$$
\be \label{3feq}
\left\{B'+ C'\cos\left[\omega\log \tau +
\frac{\Delta_\omega}{2\alpha}\ln\left(1+\left(\frac{\tau}
{\Delta_t}\right)^{2\alpha}\right)+
\frac{\Delta_\omega'}{4\alpha}\ln\left(1+\left(\frac{\tau}
{\Delta_t'}\right)^{4\alpha}\right) + \phi\right]\right\}~.
\ee
Eq.(\ref{3feq}) predicts the transition from the log-frequency $\omega$ close
to $t_c$ to $\omega + \Delta_\omega$ for $\Delta_t < \tau < \Delta_t'$ and to
the log-frequency $\omega + \Delta_\omega + \Delta_\omega'$ for $\Delta_t'
< \tau$. We stress that this corresponds to an {\it approximate} description
of a log-frequency modulation and urge the reader to take the specific
functional form of eq. (\ref{3feq}) with some caution. The purpose here is
simply to parameterise the data in order to acquire a prediction potential
for the Nikkei.

\subsection{Data analysis}

\subsubsection{The Nikkei ``bearish'' behaviour starting from 1st Jan. 1990}

\begin{figure}
\begin{center}
\epsfig{file=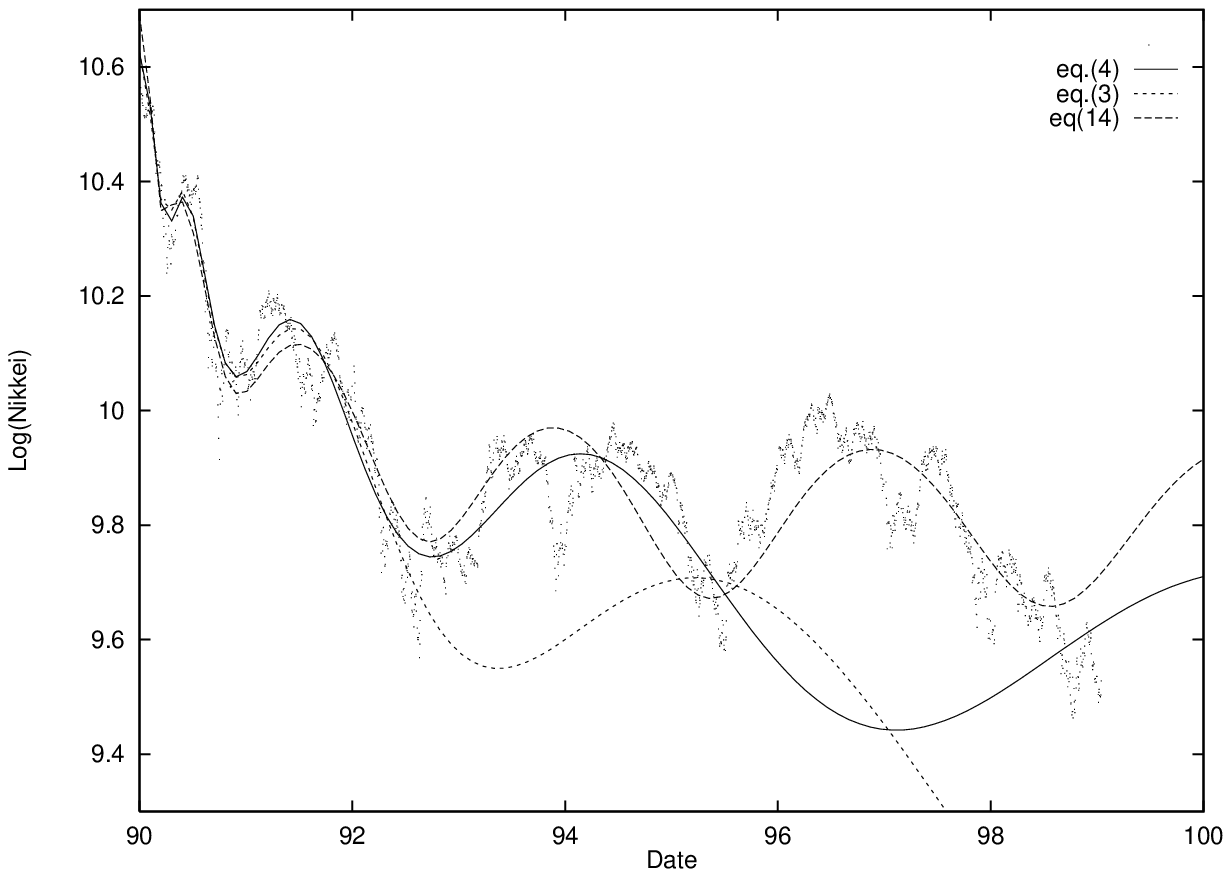,width=15cm}
\caption{\protect\label{nikkei}Natural logarithm of the Nikkei stock market
index
after the start of the decline 1. Jan 1990 until 31 dec. 1998. The lines
are eq. (\protect\ref{1feq}) (dotted line) fitted over an interval of
$\approx 2.6$
years, eq. (\protect\ref{2feq}) (continuous line) over $\approx 5.5$ years and
eq. (\protect\ref{3feq}) (dashed line) over $9$ years. The
parameter
values of the first fit of the Nikkei are $A \approx 10.7 , B\approx
-0.54 , C\approx  -0.11 , \alpha \approx  0.47 , t_c \approx  89.99 ,
\phi \approx -0.86 , \omega \approx  4.9$ for eq. (\protect\ref{1feq}).
The  parameter values of the second fit of the Nikkei are
$A' \approx
10.8 ,  B' \approx  -.70 , C' \approx  -0.11 , \alpha \approx  0.41 , t_c
\approx  89.97 ,   \phi' \approx  0.14 , \omega \approx  4.8 , \Delta_t
\approx 9.5 , \Delta_\omega \approx 4.9$ for eq. (\protect\ref{2feq}).
The third fit uses the entire time interval and is performed by adjusting only
$\Delta_t$, $\Delta_t'$,
$\Delta_\omega$ and $\Delta_\omega'$, while $\alpha$, $t_c$ and $\omega$ are
fixed at the values obtained from the previous fit. The values obtained for
these four parameters are $\Delta_t \approx 4.34$, $\Delta_t' \approx 7.83$,
$\Delta_\omega \approx -3.10 $ and $\Delta_\omega' \approx 23.4$. Note that
the values obtained for the two time scales $\Delta_t$ and $\Delta_t'$
confirms their ranking. This last fit predicts that the Nikkei should increase
as the year 2000 is approached.}
\end{center}
\end{figure}

The most recent example of a genuine long-term depression comes from Japan,
where the Nikkei has lost close to $60$ \% of its all-time high achieved on
31 Dec. 1989. In
figure \ref{nikkei}, we see the logarithm of the Nikkei from 31 Dec. 1989 until
31 dec. 1998. The fits are equations (\ref{1feq}), (\ref{2feq}) and
(\ref{3feq}) respectively with all nonlinear variables free for the two first
equations and where the interval used for the first equation is until
mid-1992 and for the second equation until mid-1995. Not only do the equations
(\ref{1feq}) and
(\ref{2feq}) agree remarkably well with respect to the parameter
values produced by the fits, but they are also in good agreement with previous
results obtained from stock market and Forex bubbles with respect to the
values of exponent $\alpha$ \cite{stockcrash2,risk}. For the fit with
(\ref{3feq}), due to the large number of free variables, we performed
differently. Of the 6 parameters $t_c$, $\alpha$, $\Delta_t$, $\omega$,
$\Delta_{\omega}$ and $\phi '$ determined from the fit with (\ref{2feq})
we kept the first 3 fixed and only $\Delta_t$, $\Delta_t'$, $\Delta_{\omega}$,
$\Delta_\omega'$ and $\phi '$ where allowed to adjust freely. The results
are given in the caption of figure \ref{nikkei}. What lends credibility to
the fit with eq. (\ref{3feq}) is that despite it complex form,
we get values for the two cross-over time scales $\Delta_t$, $\Delta_t'$
which correspond very nicely to what is expected from the Landau expansion\,:
$\Delta_t$ has moved  down to $4.4$ years, which is perfect with respect
to the interval used for the two frequency formula (\ref{2feq})
and $\Delta_t'$ is approximately $7.8$ years, which is also fully compatible with
the nine year interval of the fit. This does not mean that the
cost-function space is not very degenerate, it is, but the ranking of
$\Delta_t$ and $\Delta_t'$ is always the same and the values given does
not deviate much from the ones in the caption of figure 3, i.e., by $\pm$
one unit.

The value obtained for
$\omega \approx 4.9$ correspond to a scaling ratio $\lambda \approx 3.6$, which
is significantly larger than the $2.2 < \lambda < 2.7$ previously obtained
for the stock market \cite{risk}.
An additional difference between the Nikkei and previous results is the
strength of the oscillations compared to the leading behaviour. For the Nikkei,
it is $\approx 20$\%, i.e., 2 to 3 times as large as the amplitude
$\approx 5-10$\%
previously obtained for the stock market and the Forex \cite{risk}. We note
that the fit with eq. (\ref{1feq}) only produced the solution shown,
whereas equations (\ref{2feq}) and (\ref{3feq}) produce multiple solutions.
The solutions of equations (\ref{2feq}) and (\ref{3feq})
shown in figure \ref{nikkei} are the best solutions found which satisfies the
criteria discussed in \cite{jls,thesis}.

\subsubsection{The gold deflation price starting mid-1980}

\begin{figure}
\begin{center}
\epsfig{file=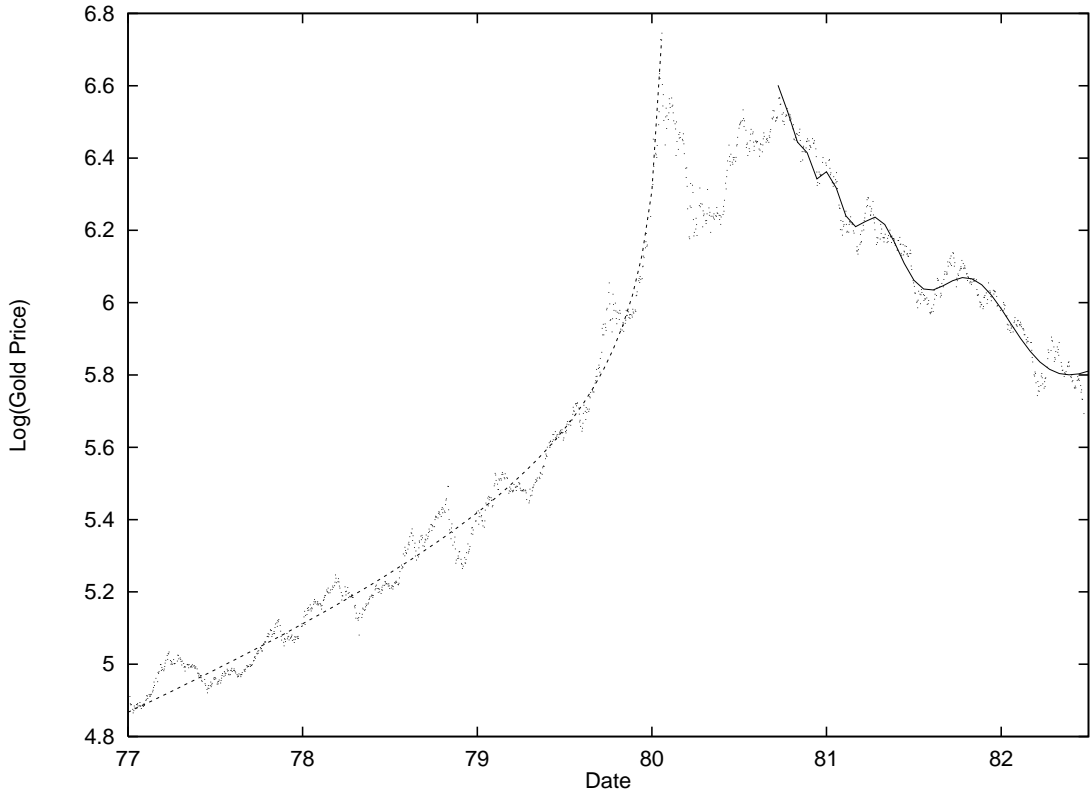,width=15cm}
\caption{\protect\label{gold} Natural logarithm of the gold 100 Oz Future
price in US \$ after the decline of the price in the early eighties. The line
after the peak is is eq. (\protect\ref{1feq}) fitted over an interval of
$\approx 2$ years. The parameter values of the fit are $A\approx 6.7 ,
B\approx  -0.69 , C\approx  0.06 , \alpha \approx  0.45 , t_c \approx  80.69 ,
\phi \approx  1.4 , \omega \approx  9.8$. The line before the peak is
 eq. (\protect\ref{1feq}) fitted over an interval of $\approx 3$ years.
The parameter values of the fit are $A\approx 8.5 ,
B \approx  -111 , C\approx  -110 , \alpha \approx  0.41 , t_c \approx  80.08 ,
\phi \approx  -3.0 , \omega \approx  0.05$}
\end{center}
\end{figure}

Another example of log-periodic decay is that of the Gold price after the
burst of the  bubble in 1980 as shown in figure \ref{gold}. The bubble has
an {\it average} power law acceleration as shown in the figure but
{\it without} any visible log-periodic structure\footnote{A pure power law
fit will not lock in on the true date of the crash, but insists on an earlier
date than the last data point. This suggests that the  behaviour of the price
might be different in some sense in the last few weeks prior to the burst of
the bubble.}.
Again, we obtain a reasonable agreement with previous results for the exponent
$\alpha$ but the value obtained for $\lambda \approx 1.9$ is now slightly
lower than
previously obtained. The strength of the oscillations compared to the leading
behaviour is again $\approx 10$\% as previously found. The fit with eq.
(\ref{1feq}) only produced the solution shown as for the Nikkei.

\subsubsection{Non-parametric power spectrum analysis}

In order to qualify further the significance of the log-periodic oscillations
in a non-parametric way, we have eliminated the leading trend from the price
data by the following transformation
\be \label{residue}
p\lp t\rp \rightarrow \frac{p\lp t\rp - \lp A +
B \tau ^{\alpha}\rp}{C\tau^{\alpha}} .
\ee

In figure \ref{decaures}, we plot the remaining residue for the case of gold.
We then analyse this residual structure by using a so-called Lomb
periodogram, which is nothing but a power spectrum analysis using a series
of local fits of a cosine (with
a phase) with some user chosen range of frequencies. The advantage of the
Lomb periodogram over a Fourier transform is that the points does not have
to be equidistantly sampled. Applying this standard technique
\cite{NumericalRecipe} to the two data sets shown in figures
\ref{nikkei}-\ref{decaures},
we construct in figure \ref{lomb} the power spectrum as a function of the
log-frequency
$f \equiv \omega/2\pi$ and find a peak at $f \approx 0.82$ for the Nikkei
and $f \approx 1.59$ for
the gold in good agreement with the fits of eq.s (\protect\ref{1feq})
and (\protect\ref{2feq}). The value $f=0.82$ obtained from the
periodogram of the Nikkei corresponds to $\lambda = 3.4$, which is
significantly higher than previously obtained for crashes.

Since the nature of the ``noise'' is unknown, we
cannot estimate the confidence interval of the peak in the standard way
\cite{NumericalRecipe}, but the relative
level of the peak for each periodogram should be regarded a measure of the
significance of the oscillations.

\section{A simple dynamical model of ``bearish'' herding behaviour}

We start from the formulation introduced in Ref.\cite{Farmer} relating the
price variation
to the net order size $\Omega$ over all traders, through a market impact
function.
Assuming that
the ratio ${\tilde p}/p$ of the price ${\tilde p}$ at which the orders are
executed
over the previous quoted price $p$ is solely a function of $\Omega$
and using the condition
that it is not possible to make profits by repeatedly trading through a close
circuit (i.e. by buying and selling with final net position equal to zero),
Farmer has
shown that the market impact function is an exponential
\be
{{\tilde p} \over p} = e^{\Omega \over L}~,
\label{qllqjdnqm}
\ee
where $L$ is the liquidity of the market. This equation is very
intuitive\,: if $\Omega >0$
(demand is larger than supply), the price increases. The reverse occurs if
supply is larger
than demand. The exponential dependence is interesting because it predicts
that the
time evolution of the logarithm of the price is given by the following equation
\be
{d \ln p \over dt} = {1 \over L} \Omega(t)~,
\label{mqmqmmq}
\ee
where the net order size $\Omega$ over all traders is changing as a
function of time so as
to reflect the information flow in the market and the evolution of the
traders' opinions and
moods.

The simplest model is that the net order size $\Omega$ is a random
uncorrelated
noise. This is the limit
where traders are heterogeneous, uncorrelated and the information flow
occurs stochastically
with no memory. This retrieves the celebrated random walk description,
since the solution of
(\ref{mqmqmmq}) is that $\ln p$ performs a random walk.

The next level of description of imitative decreasing markets
is to assume that there is a pessimistic spirit permeating the market
such that the net order sizes $\Omega(t)$ at successive times are biased
negatively, so
that there is a trend for investors to slowly exit the market and cut down
their overall positions.
However, rather
than assuming that the resulting dynamics of $\ln p$ is that of a
(negatively) biased random walk,
we posit that the negative ``bearish'' mood appears in bursts of pessimism
interspersed
in otherwise more neutral states characterised by $\langle \Omega \rangle
\approx 0$ or even
maybe slightly positive.
Furthermore, it has often been observed that investors often behave in
response to what appears to be ``psychological'' thresholds,
like the Nikkei crossing $15.000$. We will therefore
introduce such thresholds in the dynamics, which will be found essential
for the
generation of log-periodicity.
A possible mechanism for the existence of threshold and irreversible price
behaviour
is that, in a depression or in a negatively oriented market, cash
available for investment progressively disappears to pay debt or in favour
of other markets and, once
realized from a sell, may no longer be available for continuing
investment.
We model this behaviour by introducing a quenched structure in the behaviour
of the net
order size. Finally, we note that prices are given in point units
($0.01\%$) and are thus discrete.
We can thus see the log-price variation as the evolution of a walker
jumping from site to site along the axis measuring its level,
where the mesh size correspond to one point.

The transitions from one log-price to another are performed according to
the two possible
scenarios\,:
\begin{enumerate}
\item with probability $P$, the log-price surely decreases to its neighbour
below him. Furthermore,
if this occurs, the price is unable to recover its previous level and
remains for ever below.
This is the quenched structure of the model taken as a naive approximation
of the threshold
effects in financial markets.
\item with probability $1-P$, the log-price can either decrease with rate
$u$ or increase
with rate $v$.
\end{enumerate}

Thus formulated, the model is fully equivalent to the model studied by
Bernasconi and
Schneider \cite{BS} of a {\it frozen} random
lattice constructed by choosing a given configuration of randomly
distributed mixtures of the
two bond species according to their respective average concentration $P$
and $1-P$.
The exact solution of this problem has been given in \cite{BS} and shows
very clearly
nice log-periodic oscillations in the dependence of $\langle [\ln p]^2
\rangle$ as a function
of time. A simple scaling argument has been shown \cite{Review,SSS2}
to recover, in the limit $P \to 1$,
the exact results according to which the prefered log-periodic
ratio $\lambda$ is given by
\be
\lambda = \log(v/u)
\label{cjqldjq}
\ee
 and the local exponent is
\be
\alpha = \log(1-P)/ \log(u/v)~.
\ee

The mechanism of log-periodicity in this model is the result
of the interplay
between the discreteness of the log-price increments and the existence of rare
occurrence of time duration $\delta_t$
of optimisms in an otherwise ``bearish'' market, which occur with a
probability
exponentially small in $\delta_t$ as $(1-P)^{\delta_t}$ (see \cite{Review} for
a general discussion and \cite{StauSor} for another application in strongly
biased random walks in 3-d percolating networks).

This model provides a prediction that can be tested by comparing empirically
determined values of $\lambda$ and psychological moods of traders in the manner
investigated by Shiller \cite{Shiller}. Indeed, expression (\ref{cjqldjq})
predicts that a larger $\lambda$ corresponds to a larger ratio $v/u$.
In words, if $\lambda$ is larger from a devaluation than for a valuation
bubble, this means that the relative strength of the optimistic phase in
an otherwise pessimistic bearish market is stronger than the occasional
pessimism in an otherwise speculative bullish bubble. (The reverse hold for
smaller $\lambda$'s.) The study be Shiller et al \cite{Shiller1} confirm this
and show that Japanese investors are considerable more optimistic about
future trends in a decreasing Japanese stock market than in an increasing
American stock market.

\section{Conclusion}

\begin{figure}
\begin{center}
\epsfig{file=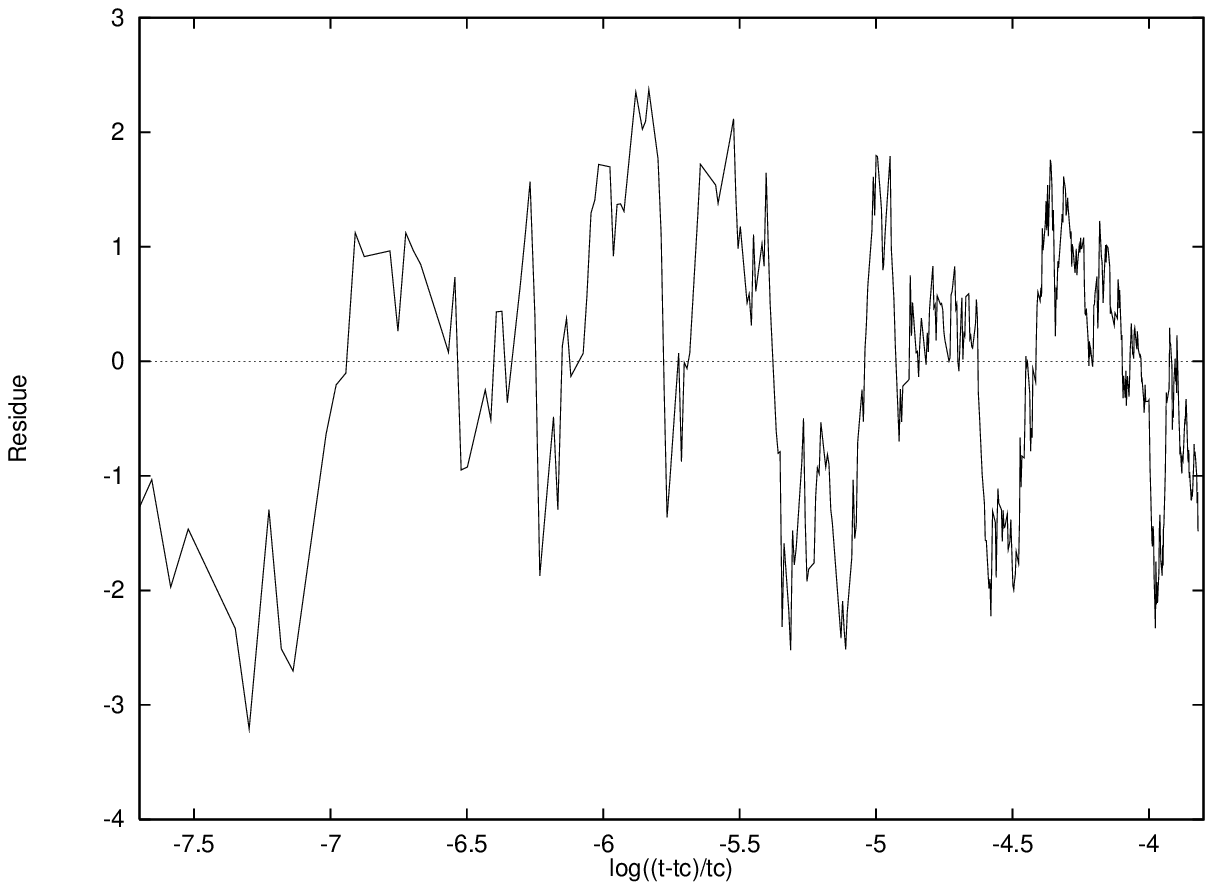,  width=10cm}
\caption{\protect\label{decaures} The residue as defined by the transformation
(\protect\ref{residue}) as a function of $\log\lp \frac{t - t_c}{t_c}\rp$ for
the 1980 gold crash.}

\vspace{1cm}

\epsfig{file=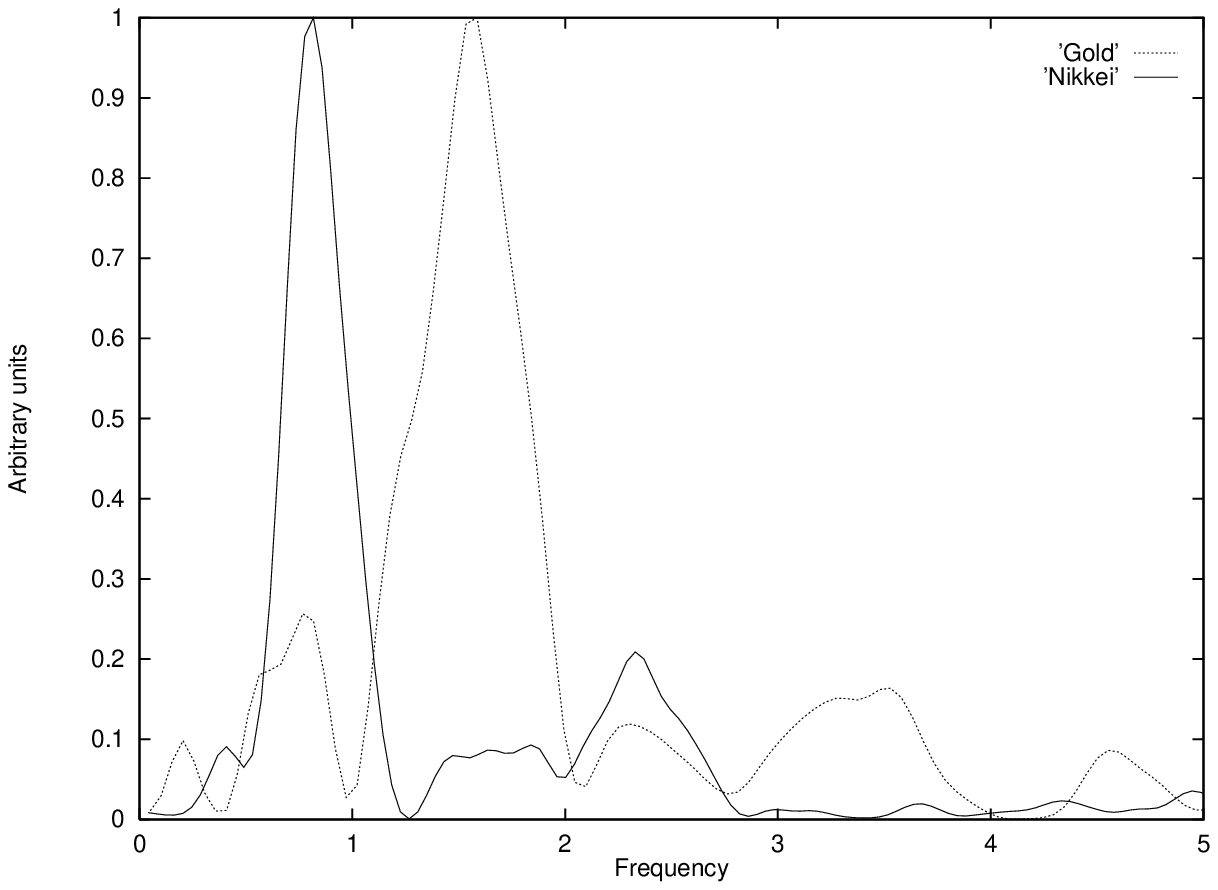=,width=10cm}
\caption{\protect\label{lomb} Lomb periodogram of the residue of the
log(price) as definded by eq. (\protect\ref{residue}) (as shown in figure
\protect\ref{decaures} for gold) of  the data shown in figures
\protect\ref{nikkei} and  \protect\ref{gold}. For each periodogram, the
significance of the peak should be estimated against the noise level.}

\end{center}
\end{figure}

We have suggested that the analogy between financial crashes and critical
phenomena can be extended to include a qualitative ensemble symmetry
with respect to the time direction. In terms of the
behaviour of the financial markets, this implies that power law decay of
prices possibly decorated by \LP \ can be found. We find this indeed to be the
case for the Nikkei stock market index and the price of gold futures. In both
cases, a good agreement is obtained between the values of the power law
exponent
for the decay of the two fits and the values previously found for the
power law increase in the market prices prior to large financial crashes.
Furthermore, the
strength of the oscillations compared to the leading behaviour indicates that
they cannot be neglected. Indeed, the Nikkei case provides us with the
strongest log-periodic signature identified in financial markets to this
date, as
clearly illustrated by the frequency analysis presented in figure \ref{lomb}
(compare the size of the peak to the background level).

However, the proposed time-symmetry operates only on an ensemble level, i.e.,
nowhere have we found log-periodicity for an increasing market followed by a
decreasing market around the same critical time $t_c$\footnote{Although
the case of gold comes ``close''.}. Furthermore,
the values found for log-frequencies and the corresponding prefered scale
ratios $\lambda$
are found less universal than for speculative bubbles ending in crashes.
That $\lambda$ differs for the Nikkei and the gold is not surprising. Gold
plays
a rather special role as a refuge in times of economic recession and strong
pessimism, which
was partly the reason for its bubble of the late seventies. Hence, the
decay of the gold price in the early eighties does not coincide with
a period of economic depression, but rather to the opposite where this
refuge is abandoned in favour of more profitable investments in bullish times.
In contrast, the decay of the Nikkei is the signature of an overall Japanese
recession. It is thus not surprising that a market with a single commodity
with such special characteristics displays more rapid oscillations with
$\lambda \approx 1.9$ than that of a global stock market.

The results published in \cite{risk} for the 6 stock market and Forex bubbles
indicates $\lambda \approx 2.5 \pm 0.3$ with the exception of the bubble
of the US\$ against the CHF, which gave $\lambda \approx 3.4$. However, the
log-periodic
signatures in this case were weaker than in the other 5 cases, i.e.,
only of $\approx 6$ \% of the leading behaviour. Hence, the larger value
presumably results from the uncertainties in the parameter estimation.

In contrast, the log-periodic signatures seen for the decay of the Nikkei
are the strongest obtained for the stock market as well as  the Forex.
This indicates that the larger value $\lambda \approx 3.4-3.6$ is not
accidental and
suggests that depressions belong to a different universality class for
log-periodicity. This is maybe not surprising
to find oscillations of a lower log-frequency
for a depression than for a speculative bubble\,: according to our model, a
larger $\lambda$ corresponds to a larger ratio $v/u$, which means that
the relative strength of the optimistic phase in an otherwise pessimistic
bearish market is
stronger than the occasional pessimism in an otherwise speculative bullish
bubble. The large $\lambda$ found for the depressed Nikkei
then reflect the fundamental psychological asymmetry
between optimism and pessimism.

\pagebreak

\newpage

\end{document}